\documentclass[conference]{IEEEtran}
\IEEEoverridecommandlockouts
\usepackage{cite}
\usepackage{amsmath,amssymb,amsfonts}
\usepackage{epsfig}
\usepackage{multirow}
\usepackage{bbding}
\usepackage{algorithmic}
\usepackage{graphicx}
\usepackage{textcomp}
\usepackage{xcolor, color}
\def\BibTeX{{\rm B\kern-.05em{\sc i\kern-.025em b}\kern-.08em
    T\kern-.1667em\lower.7ex\hbox{E}\kern-.125emX}}
\begin{document}

\title{IVCA: Inter-relation-aware Video Complexity Analyzer}

\author{
\IEEEauthorblockN{Junqi Liao$^{\dagger}$, Yao Li$^{\dagger}$, Zhuoyuan Li, Li Li$^{\star}$, Dong Liu}
\IEEEauthorblockA{University of Science and Technology of China, Hefei, China}
\IEEEauthorblockA{\{liaojq, mrliyao, zhuoyuanli\}@mail.ustc.edu.cn, \{lil1, dongeliu\}@ustc.edu.cn}\vspace{-1em}
\thanks{$\dagger$ Junqi Liao and Yao Li contribute equally to this work.}
\thanks{$\star$ Corresponding author: Li Li.}    
\thanks{This work was supported in part by the Natural Science Foundation of China under Grants 62171429 and 62021001.}
}

\maketitle

\begin{abstract}
To address the real-time analysis requirements of video streaming applications, we propose an innovative inter-relation-aware video complexity analyzer (IVCA) to enhance the existing video complexity analyzer (VCA). The IVCA overcomes the limitations of the VCA by incorporating inter-frame relations, focusing on inter motion and reference structure. To begin with, we improve the accuracy of temporal features by integrating feature-domain motion estimation into the IVCA framework, which allows for a more nuanced understanding of motion across frames. Furthermore, inspired by the hierarchical reference structures utilized in modern codecs, we introduce layer-aware weights that effectively adjust the contributions of frame complexity across different layers, ensuring a more balanced representation of video characteristics. In addition, we broaden the analysis of temporal features by considering reference frames rather than relying solely on the preceding frame, thereby enriching the contextual understanding of video content. Experimental results demonstrate a significant enhancement in complexity estimation accuracy achieved by the IVCA, coupled with a negligible increase in time complexity, indicating its potential for real-time applications in video streaming scenarios. This advancement not only improves video processing efficiency but also paves the way for more sophisticated analytical tools in video technology.
\end{abstract}


\section{Introduction}
Video data has grown explosively in recent years as an important information carrier in communication. Given the growing demand for video content, optimizing encoding parameters for videos with different content complexity is essential to ensure seamless and high-quality video streaming. A practical approach is to extract relevant complexity features to perform complexity estimation and adjust the encoding parameters.

There are two complexity estimation approaches: coding-result-based and feature-based. The coding-result-based complexity estimation methods usually establish a rate-distortion model and estimate the model parameters based on the coding results \cite{li2016lambda, liao2024content, haseeb2012rate}. Therefore, they achieve high accuracy but require high computational complexity, unsuitable for real-time scenarios. The feature-based complexity estimation methods estimate complexity based on video's spatial and temporal features \cite{menon2022vca, menon2024video, menon2023green, amirpour2024evca}, with lower complexity but challenges for accuracy improvement.

Feature-based methods are popular for real-time scenarios. ITU-T recommendations \cite{itu1999subjective} propose using spatial perceptual information (SI) and temporal perceptual information (TI) scores to assess spatial and temporal complexity. The Video Complexity Analyzer (VCA) \cite{menon2022vca, amirpour2024evca} achieves a good balance between accuracy and complexity by extracting average texture energy ($E$) and an average gradient of texture energy ($h$) as complexity features. However, VCA does not consider motion and reference structure in complexity estimation, which are the inter relation of video. From the vast difference between P-frame and I-frame coding efficiency in video coding, the inter relation of video is an essential consideration in complexity estimation.

Considering inter relation, this paper introduces the inter-relation-aware video complexity analyzer (IVCA), which is built upon the foundation of the existing video complexity analyzer (VCA). The IVCA enhances the process of complexity estimation by integrating considerations of motion and reference structure. To achieve this, we present feature-domain motion estimation (ME), which significantly improves the accuracy of temporal features by providing a more precise understanding of motion across frames. In addition, we develop layer-aware weights that effectively capture the variations in frame complexity across different hierarchical layers, taking into account the nuances of the reference structure. Furthermore, we calculate temporal features based on the reference structure instead of assuming a simple reference structure. Experimental results indicate that the IVCA not only achieves a marked improvement in accuracy compared to the VCA but does so with a negligible increase in time complexity, making it a viable option for real-time applications. Our contributions can be summarized as follows:
\begin{itemize}
    \item We design a feature-domain motion estimation to consider the influence of inter-frame motion on the temporal complexity.\vspace{0.3em}
    \item We propose a layer-aware weights scheme considering the hierarchical reference structure.\vspace{0.3em}
    \item We propose a reference-based temporal feature using the reference frame instead of the previous frame to calculate the temporal feature.\vspace{0.3em}
    \item As shown in experimental results, compared to VCA, our IVCA significantly improves complexity estimation accuracy, with negligible time complexity increase.
\end{itemize}

\begin{figure*}[t]
\centering
\centerline{\epsfig{figure=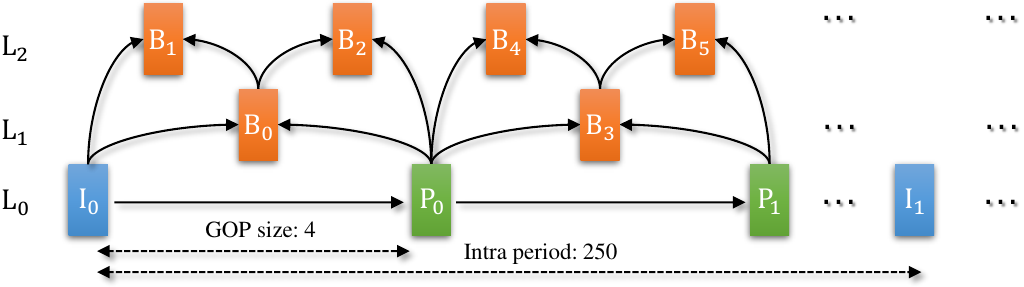,width=14.3cm}}
\caption{The reference structure in x264: 3 Layers, GOP Size 4, Intra Period 250.}
\label{fig:refer}
\vspace{-0.5em}
\end{figure*}

\section{Background}
IVCA builds upon VCA, a complexity analyzer known for its accuracy and low time complexity. To provide context for IVCA, we briefly review VCA. In VCA, a DCT-based energy function is used to evaluate the block-wise texture of each frame, defined as follows:
\begin{equation}
H_{p,k}=\sum^{w-1}_{i=0}\sum^{w-1}_{j=0}{e^{\left| \left( \frac{ij}{w^2} \right)^2-1 \right|}\left| DCT\left( i,j \right) \right|},
\end{equation}
where $k$ represents the block index in the $p^{th}$ frame, $w\times w$ denotes the block size, and $DCT(i,j)$ represents the DCT component at position $(i,j)$. Based on the calculated energy, the spatial feature $E$ can be calculated as:
\begin{equation}
E=\sum_{k=0}^{B-1}\frac{H_{p,k}}{C\cdot w^2},
\end{equation}
where $B$ represents the number of blocks per frame. The temporal feature $h$ can be calculated as:
\begin{equation}
\label{eq:VCAtemp}
h=\sum_{k=0}^{B-1}\frac{SAD\left( H_{p,k},H_{p-1,k} \right)}{C\cdot w^2},
\end{equation}
where $SAD()$ is the Sum of Absolute Differences measure. Then, the frame-level complexity $C$ is calculated by:
\begin{equation}
\label{eq:VCAseq}
C=\sum_{i=0}^{N-1}{h_i}+\sum_{j=0}^{M-1}{E_j},
\end{equation}
where $h_i$ and $E_j$ are the temporal feature of frame $i$ and spatial feature of frame $j$, respectively. $N$ and $M$ are the number of inter-coded frames and intra-coded frames, respectively. The design and implementation of VCA is straightforward, and it has proven to be very efficient in video complexity analysis~\cite{menon2022vca}. However, from the principle of VCA, it can be seen that it does not consider inter relations, including inter motion and reference relations. The inter relation is the key to affecting the efficiency of video coding. If the inter relation is not considered, the estimated complexity will not match the coding bitrate.

\section{Methods}
\subsection{Feature-domain motion estimation}
Inter-prediction is an important module in video compression which applies global or local motion estimation and compensation (MEMC) to \cite{seferidis1993general, li2017efficient, li2024uniformly, li2022global, li2024object} reduce the temporal redundancy between adjacent frames and could significantly reduce the coding bits. 
Incorporating motion information into video complexity analysis is crucial for the accuracy, particularly by excluding simple motions, as they can be easily handled by the codec with few side information bits and has limited contribution to the coding bitrate.
Based on this insight, we propose a SAD feature-domain motion estimation method to refine temporal features.

\begin{figure}[t]
\centering
\centerline{\epsfig{figure=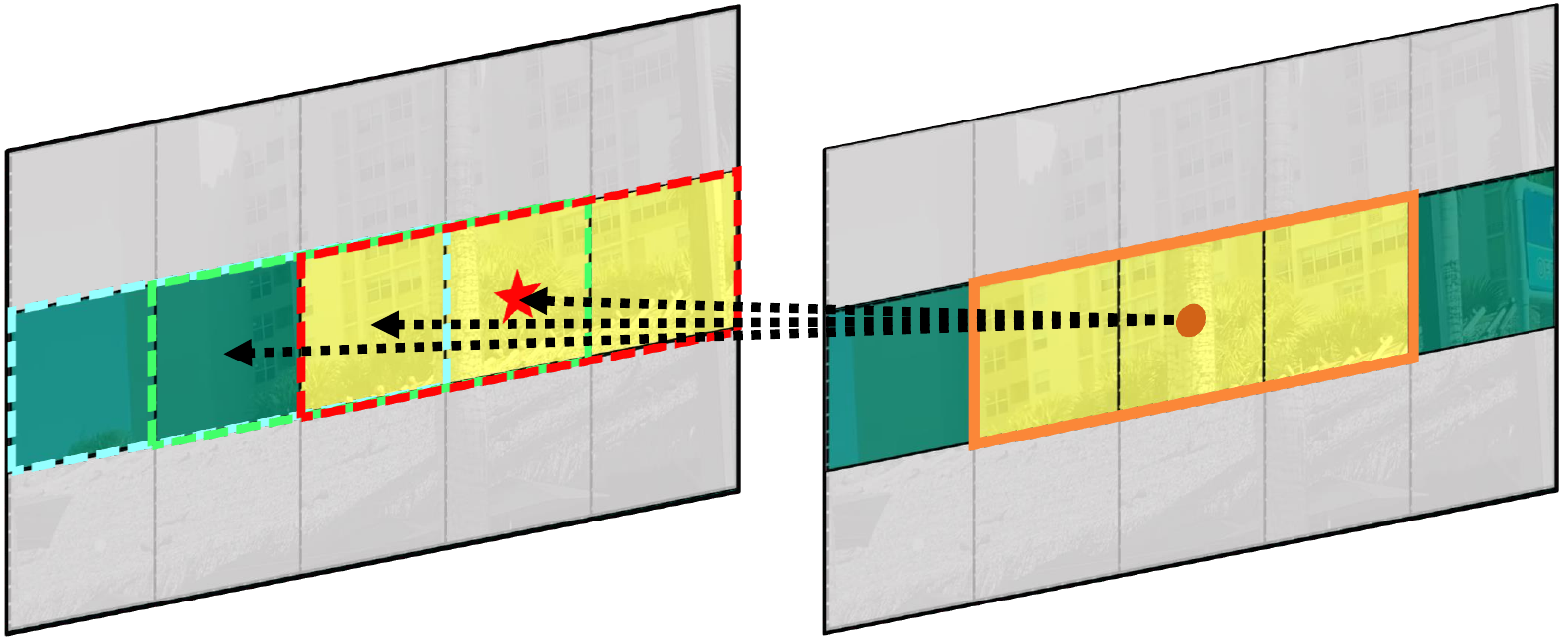,width=8.3cm}}
\caption{Illustration of the proposed feature-domain motion estimation in the horizontal direction. Blocks marked with different colors represent feature samples with different energy.}
\vspace{-1em}
\label{fig-ME}
\end{figure}

As illustrated in Fig.\ref{fig-ME}, a maximum horizontal feature cosine similarity $S_{hor}$ is firstly calculated between the current feature sample and a set of candidate reference feature samples, with neighboring feature samples also retrieved to perform the calculation as equation (\ref{eq:hor}), where $N$ stands for the sliding window size for cosine similarity calculation and $j$ stands for the candidate motion offset at the resolution of SAD feature map.
The vertical feature cosine similarity $S_{ver}$ has a similar form with the step size for the cumulative sum over $H$ changing to the image width at the block granularity $W$ as equation (\ref{eq:ver}).

\begin{figure*}
	\begin{equation}\label{eq:hor}
        S_{hor}\!=\!\underset{j}{\max}\left\{ \frac{\sum_{i=0}^{N-1}{H_{p,k+\frac{N}{2}-i}H_{p-1,k+\frac{N}{2}-i+j}}}{\sqrt{\sum_{i=0}^{N-1}{H_{p,k+\frac{N}{2}-i}^{2}}}\sqrt{\sum_{i=0}^{N-1}{H_{p-1,k+\frac{N}{2}-i+j}^{2}}}} \right\}
    \end{equation}
    \begin{equation}\label{eq:ver}
        S_{ver}\!=\!\underset{j}{\max}\left\{ \frac{\sum_{i=0}^{N-1}{H_{p,k+\left( \frac{N}{2}-i \right) W}H_{p-1,k+\left( \frac{N}{2}-i+j \right) W}}}{\sqrt{\sum_{i=0}^{N-1}{H_{p,k+\left( \frac{N}{2}-i \right) W}^{2}}}\sqrt{\sum_{i=0}^{N-1}{H_{p-1,k+\left( \frac{N}{2}-i+j \right) W}^{2}}}} \right\} 
    \end{equation}

	{\noindent} \rule[-10pt]{18cm}{0.05em}
\end{figure*}

$S_{hor}$ and $S_{ver}$ are then jointly utilized to determine the attenuation factor multiplied to temporal complexity feature $h$:
\begin{equation}
\mu =\begin{cases}
	1-\left( S_{hor}+S_{ver} \right) & \left( S_{hor}+S_{ver} \right) \leq 1\\
	1-\max \left( S_{hor},S_{ver} \right) & otherwise\\
\end{cases}.
\end{equation}
After that, the temporal complexity feature $h$ can be obtained by:\vspace{0.2em}
\begin{equation}\vspace{0.2em}
h=\sum_{k=0}^{B-1}\frac{\mu\times SAD\left( H_{p,k},H_{p-1,k} \right)}{C\cdot w^2}.
\end{equation}
$H$ can be quantized and $S^{2}_{hor}$, $S^{2}_{ver}$ may be actually used in implementation to reduce the computation complexity.

\vspace{0.2em}
\subsection{Layer-aware weights scheme}
The hierarchical reference structure ensures stable video frame quality and sequential rate-distortion performance, so it is widely used in video encoders such as x264 (as shown in Fig.\ref{fig:refer}). In a hierarchical reference structure, frames located at different layers tend to have different importance. The frame at the lower layer is referred to by more frames and is generally more critical to the sequential rate-distortion performance. Therefore, video codecs always design hierarchical quality structures corresponding to hierarchical reference structures. Inspired by this, we propose a layer-aware weights scheme to assign varying weights to frames in different layers. Furthermore, since I-frames do not refer to other frames and are encoded differently from other frames, I-frames are treated as a unique layer and are given special weights. This approach enhances the calculation of sequence-level complexity, resulting in the following improved formula based on (\ref{eq:VCAseq}):\vspace{0.2em}
\begin{equation}\vspace{0.5em}
C\!=\!w_{L_0}\!\sum_{k=0}^{O-1}{h_k}\!+\!w_{L_1}\!\sum_{m=0}^{P-1}{h_m}\!+\!w_{L_2}\!\sum_{n=0}^{Q-1}{h_n}\!+\!w_I\!\sum_{j=0}^{M-1}{E_j},\vspace{0.3em}
\end{equation}
where $O$, $P$, and $Q$ are the number of frames in layer 0, 1, and 2, respectively. $w_{L_0}$, $w_{L_1}$, and $w_{L_2}$ are the weights of layer 0, 1, and 2, respectively.

To obtain the weights, we randomly select 50 video sequences in Inter4K after excluding the test dataset and get the weights of all layers through grid search~\cite{liashchynskyi2019grid} on these 50 video sequences. The weights of the I-frame, layer 0, layer 1, and layer 2 are 0.11, 0.04, 0.0001, and 0.0005, respectively. Given that their importance in the reference structure is indeed decreasing in order, it is intuitive that their complexity weights should be decreasing in order.
\subsection{Reference-based temporal feature}
From equation (3), the VCA calculates the temporal feature according to the SAD of the current frame's texture and the previous frame's texture. That is, it implicitly assumes that the video frame only refers to the last frame. However, unlike assuming each frame refers to the previous frame, the actual reference structure of video frames follows a more complex hierarchical pattern. Thus, instead of comparing the current frame's DCT energy with the previous frame's DCT energy as in (\ref{eq:VCAtemp}), the temporal feature is calculated by comparing the current frame's DCT energy with the potential reference frame's DCT energy using SAD:\vspace{0.2em}
\begin{equation}\vspace{0.5em}
h=\sum_{k=0}^{C-1}\frac{SAD\left( H_{p,k},H_{q,k} \right)}{C\cdot w^2},
\end{equation}
where $q$ is the possible reference frame of frame $p$ according to the reference structure.

\vspace{0.5em}
\section{Experimental Results}
\begin{figure}[t]
\centering
\centerline{\epsfig{figure=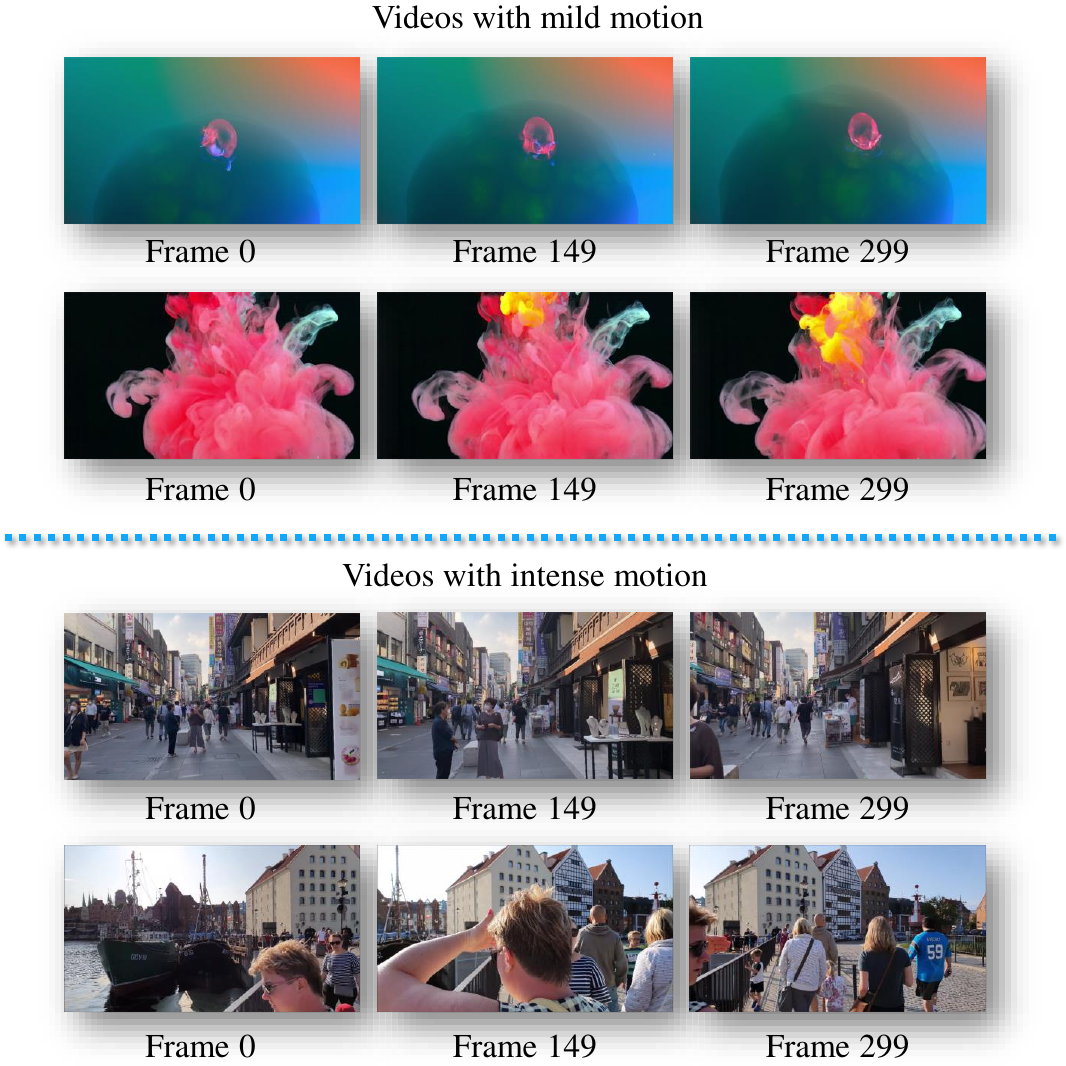,width=8.8cm}}
\caption{Examples of sequences in test dataset with mild and intense motion.}
\vspace{-0.4cm}
\label{fig-dataset}
\end{figure}

\vspace{0.2em}
The IVCA uses a comprehensive set of 50 continuous video clips sourced from the Inter4K dataset \cite{stergiou2021adapool}. As shown in Fig.\ref{fig-dataset}, this dataset is well-regarded for its diverse range of video content, providing a robust foundation for evaluating video complexity. To assess the accuracy of the IVCA, we utilize the Pearson Correlation Coefficient (PCC), which offers a statistical measure of the linear correlation between the complexity estimations and the actual coding bitrate produced by the libx264 encoder (configured with the medium preset and a Constant Rate Factor (CRF) of 26). Processing speed is measured in Frames Per Second (FPS).

\begin{table}[]
\renewcommand\arraystretch{1.5} 
\caption{Complexity estimation accuracy and speed comparison.}
\label{tabl-performance}
\centering
\scalebox{1.08}{
\begin{tabular}{ccccc}
\hline
\multicolumn{3}{l}{Schemes Applied on VCA} & \multirow{2}{*}{Accuracy}                                   & \multicolumn{1}{l}{\multirow{2}{*}{FPS}} \\ \cline{1-3}
ME       & Weighting      & Reference      &                                                             & \multicolumn{1}{l}{}                     \\ \hline
         &                &                & 79.15\%                                                     & 48.04                                    \\
\Checkmark        &                &                & 82.88\%                                                     & 48.74                                    \\
         & \Checkmark              &                & 82.08\%                                                     & 48.04                                    \\
\Checkmark        & \Checkmark              &                & \begin{tabular}[c]{@{}c@{}}86.67\% (75.70\%)\end{tabular} & 48.74                                    \\
\Checkmark        & \Checkmark              & \Checkmark              & \begin{tabular}[c]{@{}c@{}}86.42\% (76.95\%)\end{tabular} & 48.64                                    \\ \hline
\end{tabular}
}
\vspace{-0.4cm}
\end{table}

\begin{figure*}[t]
\centering
\centerline{\epsfig{figure=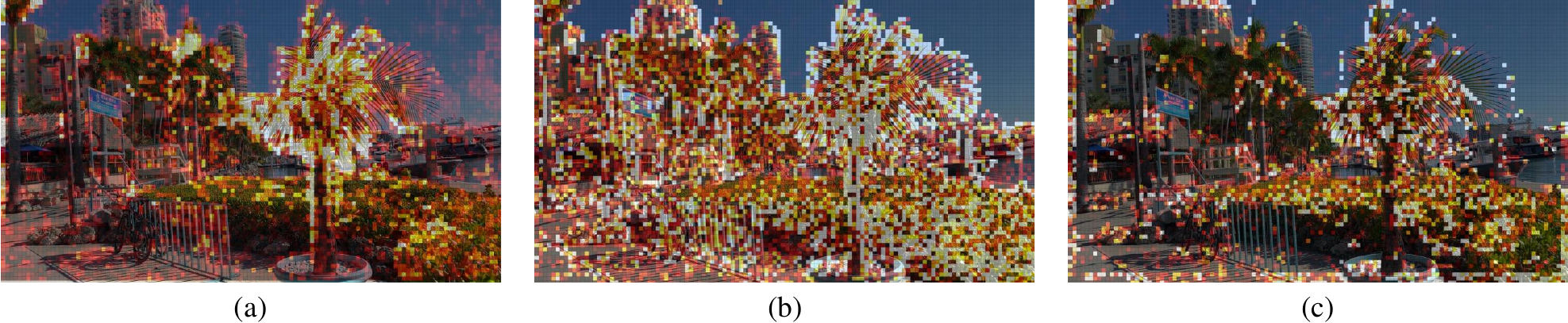,width=18cm}}
\vspace{-0.3cm}
\caption{Illustration of the distribution of libx264 coding bits and temporal complexity in VCA and IVCA on an inter-coded frame.  (a) The heat map of the libx264 coding bits. (b) The heat map of the temporal complexity in VCA. (c) The heat map of the temporal complexity in IVCA.}
\vspace{-0.2cm}
\label{fig-vca-visual}
\end{figure*}

\begin{figure}[t]
\centering
\centerline{\epsfig{figure=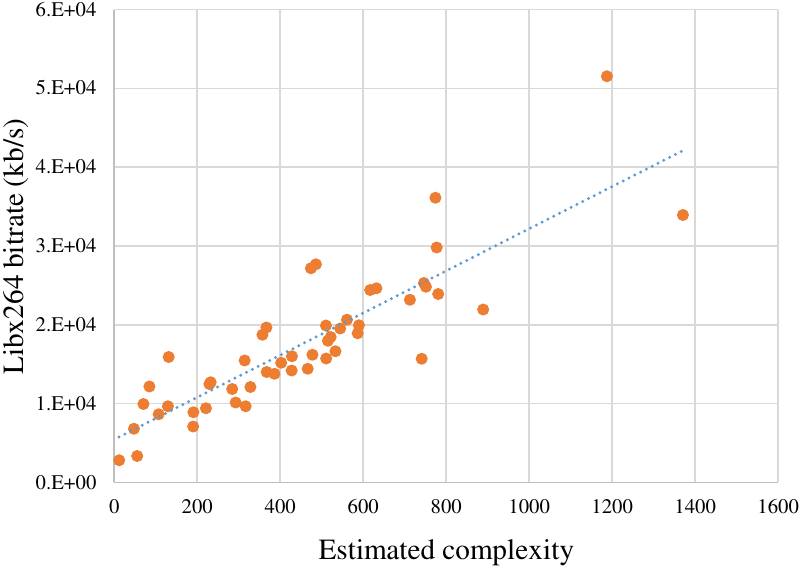,width=8.2cm}}
\vspace{-0.2cm}
\caption{A linear fitting between the IVCA estimated complexity and the actual coding bitrate of libx264.}
\vspace{-0.4cm}
\label{fig-results}
\end{figure}

Table \ref{tabl-performance} shows the effectiveness of the proposed feature-domain motion estimation method and layer-aware weights scheme, resulting in $3.73\%$ and $2.93\%$ accuracy improvements, respectively. Combining these methods achieves a total accuracy improvement of $7.52\%$. To more intuitively show the complexity estimation accuracy of our IVCA, we show the relationship between our estimated complexity on the test dataset and the actual libx264 coding bitrate in Fig.\ref{fig-results}. It can be intuitively seen from Fig.\ref{fig-results} that there is a strong positive correlation between our estimated complexity and the actual coding bitrate.
Our proposed IVCA scheme also shows a temporal complexity distribution that closely aligns with the actual coding bits of libx264 compared to VCA, as illustrated in Fig.\ref{fig-vca-visual}. Simple motion in the background is effectively handled to avoid estimation errors.

From the results obtained, the reference-based temporal feature does not enhance performance on the test dataset. This is primarily due to the mild motion in the test dataset and the insignificant differences observed among reference frames. However, as highlighted in the accompanying brackets, it exhibits superiority on a subset of the dataset with intense motion (12 video sequences in Inter4K). Consequently, when these three contributions are integrated, the IVCA performs better on datasets that feature intense motion, such as BVI-DVC \cite{ma2021bvi} and USTC-TD \cite{li2024ustc}. Furthermore, it is essential to note that the proposed IVCA exerts a negligible influence on the overall time complexity, ensuring that the efficiency of the video complexity estimation remains largely unaffected. This balance between performance enhancement and time complexity makes the IVCA particularly viable for real-time applications.

\section{Conclusion}
In this paper, we present the inter-relation-aware video complexity analyzer (IVCA) as a significant enhancement over the existing video complexity analyzer (VCA), specifically addressing its limitations that arise from a lack of consideration for inter-frame relations. By incorporating motion and reference structure, the IVCA provides a more thorough and nuanced analysis of video complexity, which is crucial for optimizing video streaming applications. The introduction of feature-domain motion estimation and the implementation of layer-aware weights and reference-based temporal features enables the IVCA to achieve markedly improved estimation accuracy. Notably, this enhancement is accomplished while maintaining a negligible increase in time complexity compared to the VCA, ensuring that the efficiency of real-time video processing is preserved. Overall, the IVCA represents a substantial advancement in video complexity analysis, paving the way for more effective and adaptive solutions in the evolving landscape of video technology.

\bibliographystyle{IEEEtran}
\bibliography{strings}

\end{document}